\documentclass[12pt,nofootinbib]{revtex}
\usepackage{amsmath}
\usepackage{amsfonts}
\usepackage{subfigure}
\usepackage{graphicx}
\usepackage{psfrag}
\usepackage{tikz}

\newcommand{\eref}[1]{Eq.~(\ref{#1})}
\newcommand{\nt}{\noindent}
\newcommand{\nn}{\nonumber}

\newcommand{\fref}[1]{Fig~\ref{#1}}
\def\refpos#1 #2 #3{\global\xrefpos=#1 \global\yrefpos=#2
                         \rlap{$\smash{#3}$}}
\def\put #1 #2 #3{\xput=#1 \yput=#2
                  \advance\xput by -\xrefpos
                  \advance\yput by -\yrefpos
                  \rlap{\kern\the\xput truebp
                        \vbox to 0pt{\vss\hbox{$\displaystyle #3$}
                        \kern\the\yput truebp}}}
\def\beginlabels\refpos#1\endlabels{\hbox{$\refpos#1$}}

\usepackage{amsfonts,amssymb}

\renewcommand{\theequation}{\arabic{section}.\arabic{equation}}

\def\slD{\slash\!\!\!\!D}

\def\be{\begin{equation}}
\def\ee{\end{equation}}

\def\bea{\begin{eqnarray}}
\def\eea{\end{eqnarray}}

\ifx\du\undefined
  \newlength{\du}
\fi
\begin{document}

\title{Holographic Non-Fermi Liquid in a Background Magnetic Field}

\author{Pallab Basu}
 \email{pallab@phas.ubc.ca}

\author{JianYang He}
\email{jyhe@phas.ubc.ca}
 
 \author{Anindya Mukherjee}
 \email{anindya@phas.ubc.ca}

\author{Hsien-Hang Shieh}
\email{shieh@phas.ubc.ca}
 
 \affiliation{Department of Physics and Astronomy, 
 University of British Columbia,
 6224 Agricultural Road,
 Vancouver, B.C. V6T 1Z1,
 Canada}

\begin{abstract}
We study the effects of a non-zero magnetic field on a class of 2+1 dimensional non-Fermi liquids, recently found in \cite{Liu:2009dm} by considering properties of a Fermionic probe in an extremal $AdS^4$ black hole background. Introducing a similar fermionic probe in a dyonic $AdS^4$ black hole geometry, we find that the effect of a magnetic field could be incorporated in a rescaling of the probe fermion's charge. From this simple fact, we observe interesting effects like gradual disappearance of the Fermi surface and quasi particle peaks at large magnetic fields and changes in other properties of the system. We also find Landau level like structures and oscillatory phenomena similar to the de Haas-van Alphen effect.
\end{abstract}

\maketitle

\section{Introduction}

Understanding the properties of strongly coupled fermionic systems is an important issue in condensed matter physics. Numerical instabilities due to the fermion sign problem and in general the complex strongly interacting nature of such problems make a numerical or an analytic study very difficult. It is believed that in some cases a non-Fermi liquid picture emerges, where a fermionic system has a sharp Fermi surface but its other properties differ significantly from those predicted from Landau's Fermi liquid theory. It is proposed that high-$T_C$ superconductors and metals close to a quantum critical point are examples of non-Fermi liquids \cite{Anderson:1990,Varma:1989}.

A general approach to study a strongly coupled system is to use the gauge/gravity duality, which may be used to map a strongly coupled field theory to a weakly coupled gravity theory \cite{Maldacena:1997re}. Many interesting condensed matter (and also fluid dynamical) phenomena have already been discussed in this context \cite{Hartnoll:2009sz,Herzog:2009xv,Rangamani:2009xk}. The systems which can be studied using gauge gravity duality are not always physically realizable. However it is hoped that studying them may provide some important insights. Recently Liu et \cite{Liu:2009dm} al have studied the simple problem of probe fermions in an extremal AdS black hole geometry\footnote{For other works on holographic non-Fermi liquids see \cite{Lee:2008xf,Cubrovic:2009ye,Rey:2008zz}.}. By studying the retarded Green's function of the fermionic field, it has been shown that at finite chemical potential a non-Fermi liquid like picture emerges. In some parameter range a well defined Fermi surface is shown to exist and  some interesting non-trivial (unlike Fermi liquid) scaling behavior is found \cite{Faulkner:2009wj}. Interestingly the scaling exponents can be calculated exactly from the scaling dimensions of the operators in a near horizon (IR) $AdS_2$ geometry.

In this paper we extend such studies to the case of non-zero magnetic fields. The study of a fermionic system in a non-zero magnetic field is by itself a vast and important subject. In this preliminary study, we are only able to explore a small part of it. We concentrate on an extremal dyonic AdS black hole geometry. Dyonic black holes are both magnetically and electrically charged and correspond to a phase of $2+1$ dimensional boundary CFT with a non-zero chemical potential and a magnetic field. Following \cite{Lee:2008xf,Liu:2009dm} we introduce a probe fermion in this geometry to study the properties of the resulting Fermi surface. We find that the introduction of magnetic field changes (decreases) the effective charge of the probe fermion. This observation enables us to address the question of the change in the nature of the Fermi surface in presence of a non-zero magnetic field. For a positive fermionic mass, the Fermi surface gets dissolved as we increase the magnetic field. We find a discrete spectrum analogous to Landau levels and discuss phenomena similar to the de Haas-van Alphen effect.

The plan of the paper is as follows, in Sec. \ref{dbh} we discuss the dyonic black hole solution and its properties. In Sec. \ref{prfe} we introduce a fermionic probe in the dyonic black hole geometry and study the resulting Dirac equations. We separate the equations into a radial and a boundary coordinate dependent part and solve the latter exactly. In Sec. \ref{res} we discuss the main results of our paper. We also include a brief review of the relevant condensed matter phenomenon in Sec. \ref{condmat}. We conclude in section Sec. \ref{conclusions} with a discussion of our results and possible future directions.
\par
\emph{Note Added:} When this work is near completion another paper \cite{Albash:2009wz} appeared in arxiv dealing with similar subject. However they did not separate their differential equations in radial and boundary coordinates and solved the full system of partial differential equations numerically. It is unclear whether a normalizability condition at special direction is implemented in \cite{Albash:2009wz} and consequently, unlike us, they do not see any discrete structure in their solution. It has been informed to us by Hong Liu that Denef, Hartnoll and Sachdev have also addressed similar subject in a recent(forthcoming) work.

\vspace{1cm}
\section{Dyonic black hole}\label{dbh}
The Maxwell-Einstein action with a negative cosmological constant in four dimensions is written as
\be
S=\frac{1}{\kappa^2}\int d^4 x \sqrt{-g}\left[R-\frac{6}{L^2}-\frac{L^2}{g_F^{2}}F_{MN}F^{MN}\right],
\ee
Here we consider the case of a dyonic black hole (i.e. with both electric and magnetic charges) in $AdS_4$ space. The metric is given by\cite{Romans:1991nq}
\begin{eqnarray}
\frac{d s^2}{L^2} &=& -f(z) d t^2+\frac{d z^2}{z^4 f(z)}+\frac{1}{z^2}(d x^2+d y^2), \\
f(z) &=& \frac{1}{z^2}\left[1+({h^2+Q^2})z^4-z^3\left(\frac1{z_+^3}+z_+(h^2+Q^2)\right)\right]
\end{eqnarray}
and the gauge fields are given by
\begin{eqnarray}
A_y=A_z=0, \hspace{1cm}A_t(z)= \mu - Qz,\hspace{1cm} A_x(y) = -hy,
\end{eqnarray}
with $\mu = Q z_+$.

\nt The Hawking temperature in this background is
\be
T_H =\frac 1{4\pi z_+}\left(3 - z_+^4 (Q^2 + h^2)\right).
\ee
\nt We may rescale the co-ordinates to set $z_+ = 1$ from now on. We wish to study the system at zero temperature i.e., in the extremal limit.  This gives us the following condition
\be
h^2  + Q ^2 =3.
\label{mag}
\ee
Note that all dimensionful quantities are now measured in units of $1/z_+$. The relevant dimensionless  parameter related to the strength of the magnetic field is $H= h/Q =\frac{h}{\sqrt{3- h ^2}} $ which goes from zero to infinity.

\section{Probe Fermion}
\label{prfe}

\nt We will consider a bulk Dirac fermion field $\Psi$ as a probe to the system. The action for $\Psi$ is given by
\begin{eqnarray}
  \label{eqn:ActionAdS4}
  S_{bulk} = \frac{1}{\kappa^2} \int d^4x
  \sqrt{-g} i \left(
      \bar{\Psi}e_{~a}^M\Gamma^a D_M\Psi -m\bar{\Psi}\Psi\right).
\end{eqnarray}
with the covariant derivative
\begin{equation}
D_M\Psi =\partial_M
\Psi+\frac{1}{8}\omega_M^{ab}[\Gamma_a,\Gamma_b]
\Psi - iqA_{M}\Psi~.
\end{equation}
Here $e_{~a}^M$ is the inverse vielbein and the non-zero components of spin-connection $\omega^{~a}_{M~b}$ are
\begin{eqnarray}
\omega_{t~z}^{~t}=-\omega_{t~t}^{~z}=\frac{1}{2}z^2 f', \hspace{1cm}
\omega_{x~x}^{~z}=-\omega_{x~z}^{~x}=\sqrt{f}=\omega_{y~y}^{~z}=-\omega_{y~z}^{~y}.
\end{eqnarray}

\nt The field $\Psi$ corresponds to a boundary fermionic operator $\mathcal{O}$. The fermion charge $q$ determines the charge of the operator $\mathcal{O}$, while its dimension $\Delta$ is determined by the mass $m$ of  $\Psi$ according to the formula
\begin{equation}
  \Delta = m + \frac32.
\end{equation}

\nt The Dirac equation $(\slD +m)\Psi=0$ can be written in the form
\begin{eqnarray}
\label{unseparated}
(U(z)+V(y))\Psi=0,
\end{eqnarray}
where
\begin{eqnarray}
U(z)&=&\Gamma^z\cdot z\sqrt{f}\left(\partial_z-\frac{1}{z}+\frac{f'}{4f}\right) +\frac{m}{z}+\Gamma^t\cdot \frac{\partial_t  - iq A_t}{z\sqrt{f}},
\nonumber \\
V(y)&=&\Gamma^x\cdot(\partial_x - iq A_x(y))+\Gamma^y\cdot \partial_y.
\end{eqnarray}
If we now perform a Fourier transform along $(t,x)$, then $\partial_t\to -i\omega,~\partial_x\to ik_x$. For simplicity, we choose the Gamma matrices as follows
\begin{equation*}
\Gamma^z = \left(
                     \begin{array}{ccc}
                     1\!\!1 & \\
                              & -1\!\!1
                      \end{array}
                      \right), \qquad
\Gamma^{\mu} = \left(
                      \begin{array}{ccc}
                         & \gamma^{\mu} \\
                         \gamma^{\mu} &
                       \end{array}
                       \right), \qquad
                       {\mathrm{with~}}
\gamma^t=i\sigma_3,~\gamma^x=\sigma_1,~\gamma^y=\sigma_2.
\end{equation*}
Then we have
\[
U=\left(\begin{array}{ccc}  D_+ & i\sigma_3 C_t \\ i\sigma_3 C_t & -D_-
\end{array}\right),  \qquad
V=\left(\begin{array}{ccc}
0 & \sigma_1 C_y+\sigma_2 \partial_y \\
\sigma_1 C_y+\sigma_2 \partial_y & 0
\end{array}\right),
\]
where
\begin{eqnarray}
D_{\pm}=z\sqrt{f}(\partial_z+C_{\pm}), &\qquad& C_{\pm}=\frac{f'}{4f}-\frac{1}{z}\pm \frac{m}{z^2\sqrt{f}}, \\
\nonumber C_t= \frac{-i\omega - iq A_t}{z\sqrt{f}}, &\qquad& C_y=i(k_x + qhy). 
\end{eqnarray}

\nt One can easily check that the matrices $U$ and $V$ do not commute. However, it is possible to find a constant matrix $M$ such that $[MU,MV]=0$. In order to separate the variables, one can left multiply \eref{unseparated} by $M$ (see \cite{Belgiorno:2008hk}). It turns out that a convenient choice for $M$ is
\[
M= \left(\begin{array}{ccc}  0& \sigma_3 \\ -\sigma_3 &0
\end{array}\right).
\]
Here
\[
MU=\left(\begin{array}{ccc}  iC_t & -\sigma_3 D_- \\  -\sigma_3 D_+ & -i C_t\end{array}\right), \hspace{1cm}
MV=\left(\begin{array}{ccc}
\sigma_3(\sigma_1C_y+\sigma_2  \partial_y) & 0 \\
0 & -\sigma_3(\sigma_1C_y+\sigma_2  \partial_y)
\end{array}\right).
\]
As $MU$ and $MV$ are two commuting hermitian matrices, we may look for solutions of the eigenvalue equation of the form
 \begin{eqnarray}
MU\Psi=-M V \Psi=L \Psi.
\end{eqnarray}
with real $L$.
Suppose
\[  \Psi=\left(\begin{array}{ccc} \Psi_+ \\ \Psi_-\end{array}\right),\]
At first we will try to solve the $y$ dependent part of the equation. It turns out that this part of the equation is identical to that of a massless free fermion in $2+1$ dimensions
\[
MV\Psi=\left( \begin{array}{ccc}\sigma_3(\sigma_1C_y+\sigma_2  \partial_y) \Psi_+ \\
-\sigma_3(\sigma_1C_y+\sigma_2  \partial_y) \Psi_-  \end{array}\right)
=-L\left(\begin{array}{ccc} \Psi_+ \\ \Psi_-\end{array}\right).
\]
 For the $\Psi_+$ part, let us left multiply $\sigma_3(\sigma_1C_y+\sigma_2 \partial_y)$, then
\begin{eqnarray}
\left[\sigma_3(\sigma_1C_y+\sigma_2  \partial_y)\right]\left[\sigma_3(\sigma_1C_y+\sigma_2  \partial_y)\right]\Psi_+=L^2 \Psi_+.
\end{eqnarray}
After simplification we get
\begin{eqnarray}
\partial_y^2\Psi_+  + (L^2+C_y^2-i\sigma_3 C'_{y})\Psi_+=0.
\end{eqnarray}
Suppose $\Psi_+^T=(R_1(z)S_1(y),R_2(z)S_2(y))$. Since the equation above does not depend on $z$, one can set $R_1=R_2$, leading to
\begin{eqnarray}
-\partial_y^2 S_{1,2} +P_{\pm}(y)S_{1,2}=0,
\end{eqnarray}
where
\begin{eqnarray}
P_{\pm}(y)=-(L^2+C_y^2\mp i C_y')=(k_x + qhy)^2-L^2 \mp qh,
\end{eqnarray}
which is exactly a simple harmonic oscillator potential. To see this more explicitly, let us define
$\eta=\sqrt{q h}(y + \frac{k_x}{qh})$. Then
\begin{eqnarray}
-\partial_{\eta}^2 S_{1,2}+\eta^2 S_{1,2}=\left(\frac{L^2}{ qh} \pm 1\right)S_{1,2}.
\end{eqnarray}
The eigenvalues and eigenvectors are
\begin{eqnarray}
E_n &=& \frac{1}{2}\left(\frac{L^2}{qh} \pm 1\right)=n+\frac{1}{2}, ~~n=0,1,2,\dots  \nonumber \\
S_{1,2} &=& N_n e^{-\eta^2/2} H_n(\eta)\equiv I_n(\eta),
\end{eqnarray}
where $N_n$ are normalized factors. Substituting back into the first order equations for $\Psi_+$ with the same eigenvalues we get
\[ \Psi_+=\left(\begin{array}{ccc} I_{n}(\eta)R_1 \\ -i I_{n-1}(\eta)R_1 \end{array}\right),\]
corresponding to eigenvalues $L_n^2= 2nqh,
~n=0,1,2,\cdots$. For completeness, we define $I_{-1}(\eta)=0$.
Combining with the expression for $\Psi_-$ which is obtained similarly, we have
\begin{eqnarray}
 \Psi=\left(\begin{array}{ccc} I_{n}(\eta)R_1 \\ -i I_{n-1}(\eta)R_1 \\ I_{n}(\eta)R_2 \\ i I_{n-1}(\eta) R_2\end{array}\right).
\end{eqnarray}
\nt Another independent solution comes from the values corresponding to $-L_n$, i.e.
\begin{eqnarray}
MU\Psi'=-MV\Psi'=-L_n\Psi',
\end{eqnarray}
with
\begin{equation}
\Psi'=\left(\begin{array}{ccc}
 -iI_{n}\bar{R}_1 \\  I_{n-1}\bar{R}_1 \\ -iI_{n}\bar{R}_2 \\ - I_{n-1} \bar{R}_2
\end{array}\right).
\end{equation}

\nt The equations for $R_i$ and $\bar{R}_i$ are respectively,
\begin{eqnarray}
D_+R_1=-(iC_t+L_n)R_2, \hspace{1cm} D_-R_2=(iC_t-L_n)R_1,  \nonumber\\
D_+\bar{R}_1=-(iC_t-L_n)\bar{R}_2, \hspace{1cm} D_-\bar{R}_2=(iC_t+L_n)\bar{R}_1,
\end{eqnarray}

\nt Note that for $n=0, L_0=0$, and the solution is
\[ \Psi=-i\Psi'=\left(\begin{array}{ccc} I_0(\eta)R_1 \\0 \\  I_0(\eta) R_2 \\0 \end{array}\right),\]
with $R_i=\bar{R}_i$. Thus in this case there is only one independent solution.

\subsection{Green's Function}

Let's first look at the general form of the solutions\footnote{Here we have used a somewhat nonconventional notation for the expansion.
We have absorbed the coefficients of the eigenfunctions into the definitions of $R_{1, 2}^{(n)}$.
Explicitly, these coefficients are specified by assigning appropriate asymptotic boundary conditions.
This is to facilitate comparisons to the Green's functions in \cite{Liu:2009dm,Faulkner:2009wj}.}
\[
\Psi=\sum_n\Psi^{(n)}, \hspace{1cm}
\Psi^{(n)}=\Psi^{(L_n)}+\Psi'^{(-L_n)}=
\left(\begin{array}{ccc} iR_1^{(n)}e_1^{(n)} \\  R_2^{(n)}e_2^{(n)} \end{array}\right)
+\left(\begin{array}{ccc} -i\bar{R}_1^{(n)}e_2^{(n)} \\  -\bar{R}_2^{(n)}e_1^{(n)} \end{array}\right),
\]
i.e.
\begin{eqnarray}
\Psi_+^{(n)}=iR_1^{(n)}e_1^{(n)} -i\bar{R}_1^{(n)}e_2^{(n)}, \hspace{1cm}
\Psi_-^{(n)}=-\bar{R}_2^{(n)}e_1^{(n)} +R_2^{(n)}e_2^{(n)}.
\end{eqnarray}
For simplicity, we have introduced a set of two-component basis spinors
\[
e_1^{(n)}=\left(\begin{array}{ccc} i I_{n}(\eta) \\  I_{n-1}(\eta) \end{array}\right), \qquad
e_2^{(n)}=\left(\begin{array}{ccc}  I_{n}(\eta) \\ i I_{n-1}(\eta) \end{array}\right), \qquad
n\geq 1.
\]
The orthonormality of the solutions to the $y$ dependent equations gives
\begin{eqnarray}
\int d\eta \Big(e_1^{(n)}\Big)^\dagger e_2^{(n')}=\int d\eta \Big(e_2^{(n)}\Big)^\dagger e_1^{(n')}=0, &\hspace{1cm}
\int d\eta \Big(e_1^{(n)}\Big)^\dagger e_1^{(n')}=\int d\eta \Big(e_2^{(n)}\Big)^\dagger e_2^{(n')}=2\delta_{nn'}, \nonumber\\
i\sigma_3 e_1^{(n)}=-e_2^{(n)},&\hspace{1cm} i\sigma_3 e_2^{(n)}=e_1^{(n)}.
\end{eqnarray}
If we express the spinors $e_i^{(n)}$ as
\[
e_1^{(n)}\Rightarrow\epsilon_1^{(n)} =\left(\begin{array}{ccc} 1 \\  0 \end{array}\right), \qquad
e_2^{(n)}\Rightarrow\epsilon_2 ^{(n)}=\left(\begin{array}{ccc} 0 \\ 1 \end{array}\right),
\]
then in this basis $i\sigma_3=\left(\begin{array}{ccc} 0 & 1 \\ -1 & 0 \end{array}\right)$.

We now have everything we need to calculate the Green's function. Given the orthonormality property introduced above, we can as usual substitute the solutions into the boundary term in the action

\begin{eqnarray}
  \label{eqn:boundary}
  S_{b} = \int d^3x
  \sqrt{-g_3} \bar{\Psi}_+\Psi_- .
\end{eqnarray}

\nt In the basis of $(\epsilon_1^{(n)},\epsilon_2^{(n)})$
\[
\Psi_+^{(n)}=\left(\begin{array}{ccc} iR_1^{(n)} \\  -i\bar{R}_1^{(n)} \end{array}\right), \hspace{1cm}
\Psi_-^{(n)}=\left(\begin{array}{ccc}  -\bar{R}_2^{(n)}\\ R_2^{(n)} \end{array}\right)
   =-i\left(\begin{array}{ccc}  0 & \frac{\bar{R}_2^{(n)}}{\bar{R}_1^{(n)}}\\ \frac{R_2^{(n)}}{R_1^{(n)}} &0\end{array}\right) \Psi_+^{(n)}
   \equiv S^{(n)}\Psi_+^{(n)}
\]
the retarded Green's function is given by \cite{Liu:2009dm},
\[
G_R^{(n)}=-i\sigma^t S^{(n)}= -i\cdot i\sigma_3\cdot(-i)
   \left(\begin{array}{ccc}  0 & \frac{\bar{R}_2^{(n)}}{\bar{R}_1^{(n)}}\\ \frac{R_2^{(n)}}{R_1^{(n)}} &0\end{array}\right)
=-\left(\begin{array}{ccc}  0 & 1\\ -1 & 0 \end{array}\right)
   \left(\begin{array}{ccc}  0 & \frac{\bar{R}_2^{(n)}}{\bar{R}_1^{(n)}}\\ \frac{R_2^{(n)}}{R_1^{(n)}} &0\end{array}\right)
=\left(\begin{array}{ccc}  -\frac{R_2^{(n)}}{R_1^{(n)}} &0\\ 0& \frac{\bar{R}_2^{(n)}}{\bar{R}_1^{(n)}} \end{array}\right).
\]
The spectral function is
\begin{eqnarray}
A^{(n)}\sim -{\mathrm{Im}} ({\mathrm{Tr}} ~i\sigma^t S^{(n)})
  \sim  {\mathrm{Im}}\left(-\frac{R^{(n)}_2}{R^{(n)}_1}+\frac{\bar{R}^{(n)}_2}{\bar{R}^{(n)}_1}\right).
\label{spect}
\end{eqnarray}

\nt The special case of $n=0$ needs to be considered separately. It is easy to show that
\[\Psi_-^{(0)}=i I_0 R_2^{(0)} \left(\begin{array}{ccc}0 \\ 1\end{array}\right)
 =-\frac{R_2^{(0)}}{R_1^{(0)}} \Psi_+^{(0)} \equiv S^{(0)}\Psi_+^{(0)},
\]
and the Green's function is
\[G_R^{(0)} =-i\sigma^t S^{(0)}=\sigma^3 S^{(0)}=-\frac{R_2^{(0)}}{R_1^{(0)}}
   \left(\begin{array}{ccc} 1 && 0 \\ 0&& -1\end{array}\right),
\]
and thus the spectral function vanishes,
\begin{eqnarray}
A^{(0)}\sim -{\mathrm{Im}} ({\mathrm{Tr}} ~i\sigma^t S^{(0)}) =0.
\end{eqnarray}

\nt Let us introduce $\xi^{(n)}=\frac{R_2^{(n)}}{R_1^{(n)}},\bar{\xi}^{(n)}=\frac{\bar{R}_2^{(n)}}{\bar{R}_1^{(n)}}$
to discuss the general behavior of Green's function.
The equations for them are
\begin{eqnarray}
z\sqrt{f}\partial_z\xi^{(n)} &=& \xi^{(n)}\frac{2m}{z}+(\xi^{(n)})^2(iC_t+L_n)+(iC_t-L_n), \nonumber\\
z\sqrt{f}\partial_z\bar{\xi}^{(n)} &=& \bar{\xi}^{(n)}\frac{2m}{z}+(\bar{\xi}^{(n)})^2(iC_t-L_n)+(iC_t+L_n),
\label{maineq}
\end{eqnarray}
which are exactly the same equations appearing in (24) of \cite{Liu:2009dm} with $\bar{\xi}^{(n)}\sim \xi_-,\xi^{(n)}\sim \xi_+$ and $L_n\sim k_1$.

The functions $\xi$ and $\bar \xi$ satisfy
\begin{eqnarray}
\bar{\xi}(L_n)\sim \xi(-L_n),
\end{eqnarray}
and for $m=0$
\begin{eqnarray}
\bar{\xi}^{(n)}=-\frac{1}{\xi^{(n)}} ~~~ \longrightarrow ~~~\xi(n=0)=\bar{\xi}(n=0)=i.
\end{eqnarray}

\subsection{Boundary Conditions at the Horizon}

\nt In order to solve for the Green's function, one needs a boundary condition at the horizon, $z\to 1$. To simplify the equations in the near horizon limit, let us introduce $\tilde{R}_i=f^{1/4}R_i$, and note that for an extremal black hole
\begin{eqnarray}
f=1+3z^4-4z^3\to 6(1-z)^2.
\end{eqnarray}
In the near horizon limit we get
\begin{eqnarray}
\partial_u^2\tilde{R}_i=-\omega^2\tilde{R}_i, \hspace{1cm}{\mathrm{with~~}}u\sim\int\frac{dz}{f(z)}\sim -\frac{1}{6(1-z)}.
\end{eqnarray}
To compute the retarded Green's function we will impose the in-falling wave condition at the horizon
\begin{eqnarray}
R_1\sim \frac{1}{\sqrt{1-z}}e^{\frac{i\omega}{6(1-z)}}, \hspace{1cm}R_2=i R_1.
\end{eqnarray}

For $\omega=0$, the above mentioned in-falling boundary conditions do not apply. Here (see \cite{Liu:2009dm}),
\begin{equation}
 \xi(\bar \xi)= \frac{m-\sqrt{k^2+m^2-\frac{(qQ)^2}{6}-i\epsilon}} {\frac{qQ}{\sqrt{6}}+(-) k}
\end{equation}

\section{Results} \label{res}

In order to understand the Fermi level structure and the associated critical behavior we plot the spectral function (\eref{spect})
as a function of $L_n$ (or $n$) and $\omega$. Due to  quantization of  the energy levels in presence of a magnetic field the quantity $L_n$ is discrete.
However, we will at first try to solve the equations \eref{maineq} as if $L_n$ is a continuous parameter. The charge of the fermion $q$ appears
in \eref{maineq} only in the form $qQ$, where $Q$ is the charge of the black hole. Hence as long as we keep $qQ$ fixed
the nature of the solution should not change. For a given magnetic field $h$ we can define $q_{eff}=q\sqrt{1-\frac{h^2}{3}}$,
such that a system with charge $q$ and magnetic field $h$ is equivalent to a system with charge $q_{eff}$ and zero magnetic field.
This can be written schematically as in \eref{rel}. This decrease in the effective charge is not surprising because
for a fixed horizon radius of the black hole, chemical potential decreases with increasing $h$ ( \eref{mag}).
\begin{equation}
 (q,h) \rightarrow (q_{eff}=q\sqrt{1-\frac{h^2}{3}},0)
\label{rel}
\end{equation}
From the above argument we can conclude that the physical properties of our system are similar to what has been calculated
in the zero magnetic field case \cite{Faulkner:2009wj}. However for a fixed probe fermion charge $q$ if we change the magnetic field,
$q_{eff}$ changes. This leads to a subtle relation between Fermilevel structure and the applied magnetic field.
Here we will not try to calculate all the properties of the system. Instead we will just briefly summarize a few pertinent aspects
using the results of \cite{Faulkner:2009wj}. For a more detailed and complete discussion we refer to the original literature.
At zero magnetic field the role of $L_n$ is played by the momentum $k$. As long as we use $L_n$ as a continuous parameter
the discreteness of the energy levels does not play any role in our discussion.
Of course, after we have a profile for the spectral function we have to put back the discrete energy levels for a physical interpretation.

\begin{figure}[htbp]
\begin{center}
\includegraphics[scale=0.3]{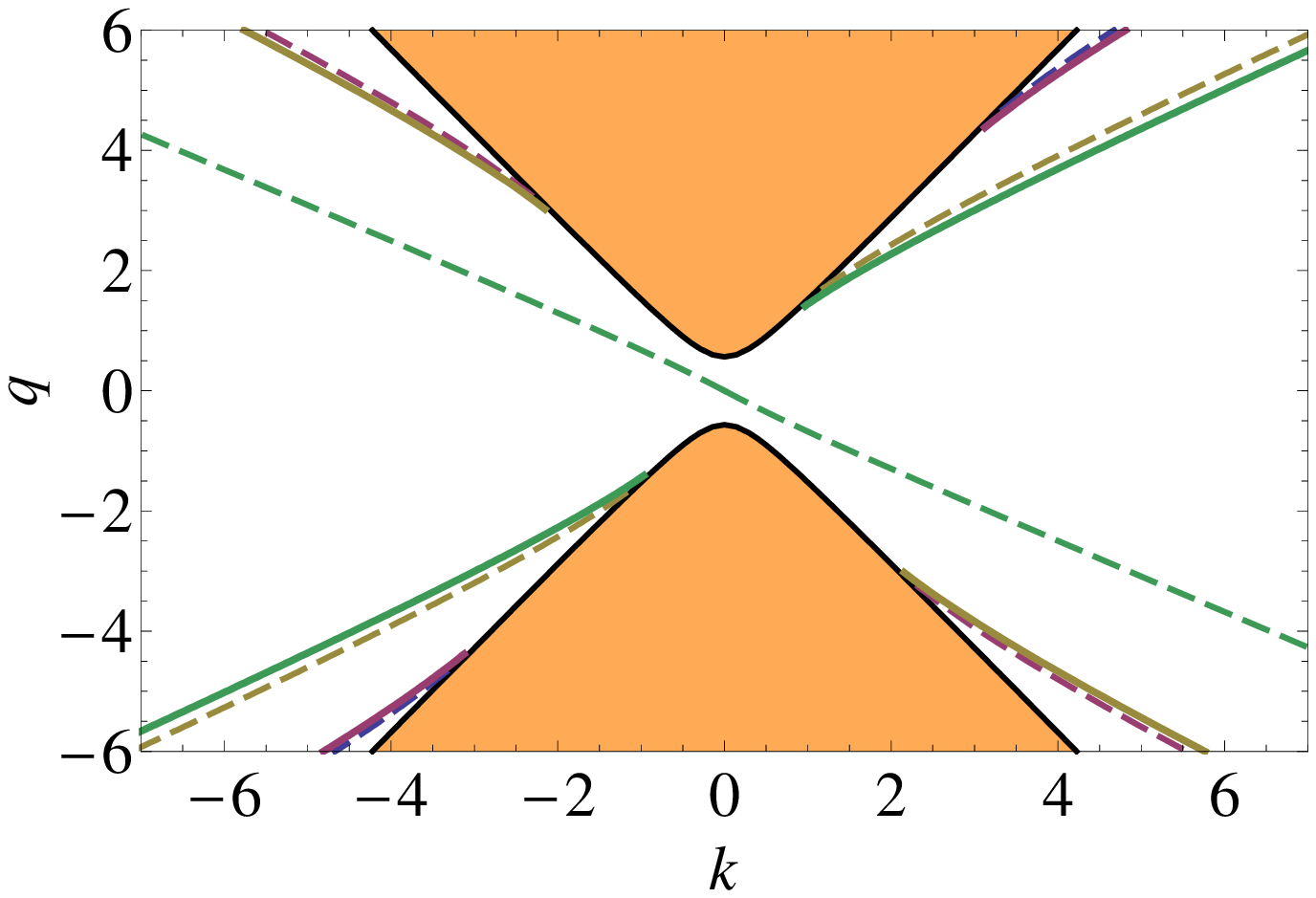}
\includegraphics[scale=0.3]{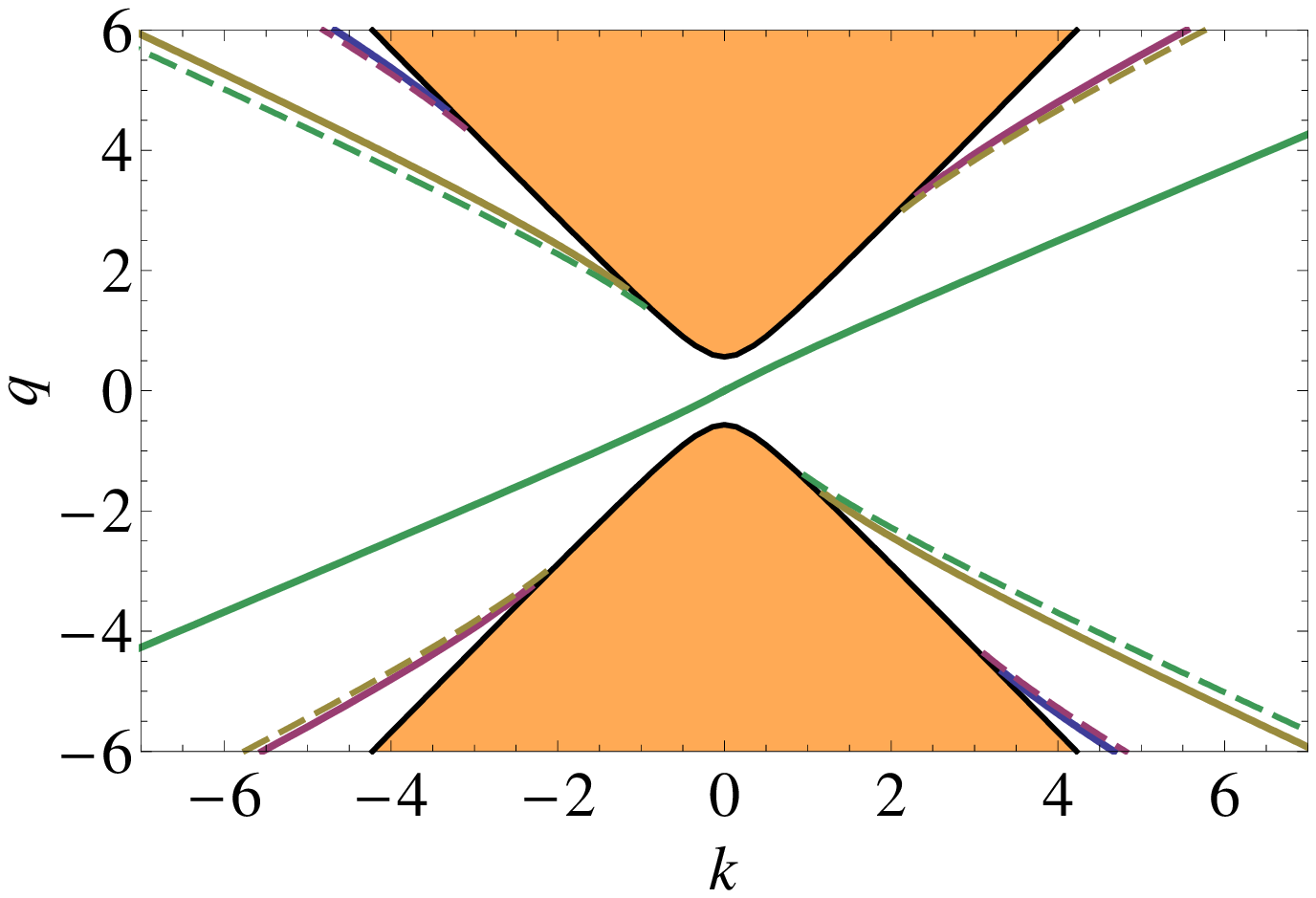}
\includegraphics[scale=0.3]{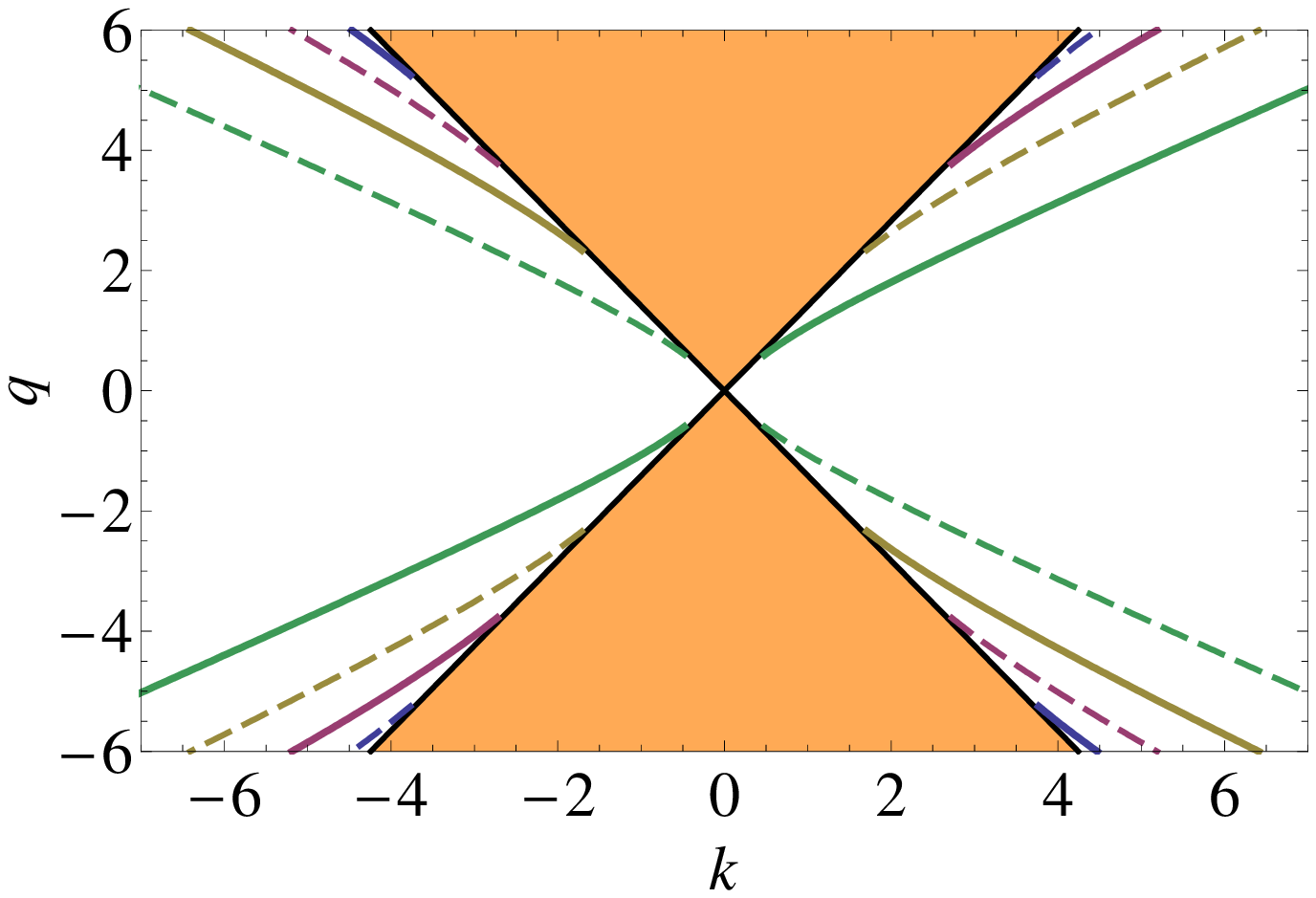}
\caption{Fig 5 of \cite{Faulkner:2009wj}. The values of $k_F$ as a function of $q$ are shown by the solid lines for $m=0.4,m=-0.4,m=0$. The oscillatory region with imaginary $\nu_k$ is shaded. As we turn on a non-zero magnetic field $q$ changes to $q_{eff}=q\sqrt{1-\frac{h^2}{3}}$. From this diagram one can easily see what happens to the system.}
\label{fig:fermi}
\end{center}
\end{figure}

\begin{itemize}
 \item The existence of Fermi surface is manifested by the existence of one or more $\omega=0$ bound states of \eref{maineq}. A bound state corresponds to a sharp peak in $\mathrm{Im}~G^{22}_R(0,k)$. At large $q$ there exists one or more bound states of \eref{maineq}. These bound states can be found numerically or by WKB approximation. In the region $\omega \rightarrow 0,k \rightarrow k_F$
\begin{equation}
 G^{22}_R(k,\omega)=\frac{h_1}{k-k_F-\frac{1}{v_F}\omega-h_2 e^{i\gamma_{k_F}}\omega^{2\nu_{k_F}}}
\end{equation}
The pole of the Green's function in the complex $\omega$ plane is located at
\begin{equation}
 \omega_c(k) = \omega_*(k)-i \Gamma(k),
\label{pole}
\end{equation}
where we have
\begin{eqnarray}
\label{critical}
 \omega_*(k) \propto (k-k_F)^z \\
 \nn \Gamma(k) \propto (k-k_F)^\delta
\end{eqnarray}
The critical exponents are given by
\begin{equation}
 z=\{\begin{array}{c}
 \frac{1}{2\nu_{k_F}}, \nu_{k_F} < \frac{1}{2} \\
1,\nu_{k_F} > \frac{1}{2}
\end{array}
\label{scaling1}
\end{equation}

and
 \begin{equation}
  \delta=\{\begin{array}{c}
 \frac{1}{2\nu_{k_F}}, \nu_{k_F} < \frac{1}{2} \\
2\nu_{k_F},\nu_{k_F} > \frac{1}{2}.
 \end{array}
\label{scaling2}
 \end{equation}
From such critical behavior it follows that the ratio $\frac{\Gamma}{\omega_*}$ goes to zero if $\nu_{k_F} > \frac{1}{2}$ and to a non-zero constant otherwise. Hence for $\nu_{k_F} < \frac{1}{2}$ the imaginary part of the pole is always comparable to the real part and thus the quasi-particle is never stable. Another important feature of the pole is that its residue goes to zero as the Fermi surface is approached. In particular, the smaller $\nu_{k_F}$ , the faster the residue approaches zero. For $\nu_k=\frac{1}{2}$, the Green's function has logarithmic corrections. $\nu_{k_F}$ can be calculated exactly from looking at the near horizon (IR) $AdS_2$ region. In the IR geometry, the conformal dimension $\delta_k$ of the operator corresponding to the probe fermion with momentum $k$ is given by
\begin{equation}
 \delta_k=\frac{1}{2}+\nu_k, \qquad \nu_k=\sqrt{\frac{k^2+m^2}{6}-\frac{q^2}{12}}.
\label{nuk}
\end{equation}

\item For a fixed $m$ there always exists an ``oscillatory region'' in the $(q,k)$ parameter space (\fref{fig:fermi}) characterized by an imaginary $\nu_k$. If $m \ge 0$ and $q$ is gradually decreased the Fermi levels dissolve in the oscillatory region (\fref{fig:fermi}).At this point there will no longer be any bound state solutions to \eref{maineq}. For $m < 0$ the Fermi level still exists for small $q$.

\item If $m \neq 0$ then decreasing $q$ leads to a phase where there is no oscillatory region (\fref{fig:fermi}). For $m=0$ however the oscillatory region persists down to $q=0$.

\item There are other notable features of the Green's function like a finite peak at $\omega \approx k-\sqrt{3} q$ etc. We refer the reader to \cite{Liu:2009dm} for a more elaborate discussion.
\end{itemize}

\nt If we start with a value of $q=q_{ini},h=0$ and begin to increase $h$ then $q_{eff}$ decreases. The evolution of the system is understood form \fref{fig:fermi}. Generically $k_F$ decreases with increasing magnetic field, it is expected that $\nu_k$ will also decrease. For $\nu_{k_F} < \frac{1}{2}$ the quasi particle becomes unstable. For $m > 0$, it seems that there exist a $h=h_{c}(m,q_{ini})$ such that $\nu_{k_F}$ becomes zero at this point.  The system enters the oscillatory region then and there is no Fermi surface for $h>h_c$. As shown in \cite{Faulkner:2009wj} using a WKB approximation, the Fermi surfaces only exist for $m^2 < q^2_{eff}/3$. Hence 
\begin{equation}
 h_c=\sqrt{3(1-3\frac{m^2}{q^2_{ini}})}
\end{equation}
It should be noted that at the cross over phase, $\nu_K\approx 0$ and the system is far from being a Fermiliquid.  The oscillatory region itself goes away for a greater value of $h$.

\begin{figure}[htbp]
\begin{center}
\includegraphics[scale=0.8]{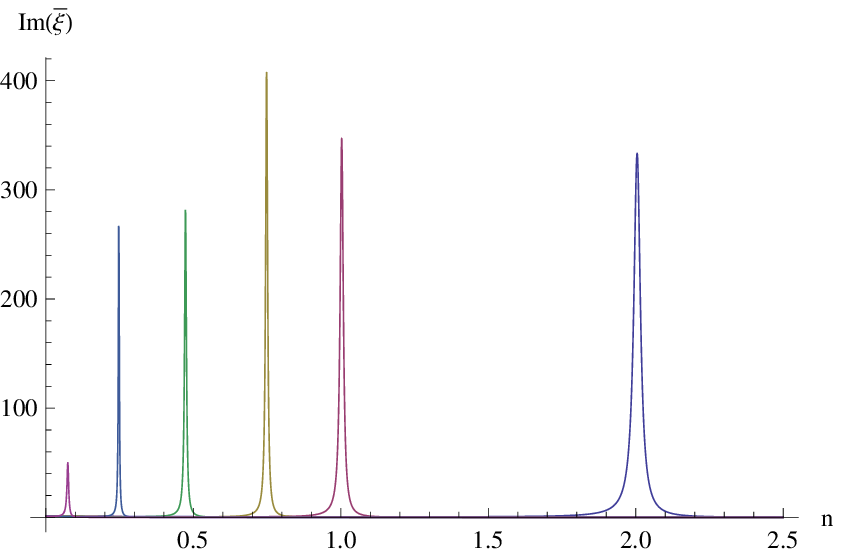}
\includegraphics[scale=0.8]{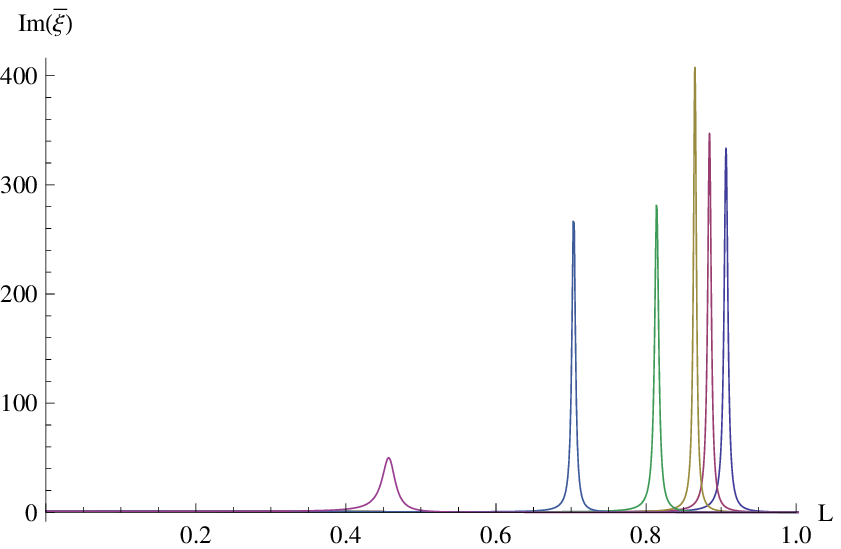}
\caption{The poles of $\mathrm{Im}~\bar{\xi}$ at $\omega=-10^{-9}$, $m=0$ with different magnetic fields $h$ and $q_{ini}=1$: from right to left, $h=0.205,~0.39,~0.5,~0.7,~1.0,~1.4$ respectively.
}
\label{fig:poles_moving}
\end{center}
\end{figure}

For the case $m \le 0$, when the magnetic field increases, the pole seems to persists for arbitrary small $q_{eff}$. Hence there will be no disappearance of Fermi surface at strong magnetic field. Here we have plotted (\fref{fig:poles_moving}) how the location of the pole (Fermi surface) moves with back ground magnetic field $h$.

\subsection{Discrete nature of $L_n$}

The above discussion is not complete. One important aspect of our solution is that $n=0,1,2,\cdots$ is discrete (and hence so is $L_n$), rather than continuous as the plane wave number $k$ in \cite{Liu:2009dm}. Due to the discrete nature of $L_n$ the entire $(\omega, L_n)$ space is not accessible. The only physically allowed values are $L_n=\sqrt{2nqh}$.\footnote{ For large $n$, the distance between consecutive levels is proportional to $\sqrt{\frac{qh}{n}}$. Hence for large $n$, the levels become densely packed. One recovers a continuum at zero magnetic field by defining a limit keeping $L_n=\sqrt{2nqh}$ fixed while $h \rightarrow 0$, $n\rightarrow \infty$. At this limit our system becomes same as the zero magnetic field case \cite{Liu:2009dm}. }  This is the analogue of Landau levels. However, discrete values of $L_n$ do not necessarily imply a discrete spectrum in the dyonic black hole background.\footnote{The holographic dual of a black hole is a strongly coupled large $N$ gauge theory at finite energy density. Hence the spectrum is most likely continuous due to interactions \cite{Festuccia:2006sa}. Our situation is not be very different from the case of a global AdS black hole, where due to the compactness of the boundary the $S^3$ harmonics are quantized, although the spectrum in a black hole background is not discrete.} One important difference from the free fermions case comes from the infalling boundary condition at the horizon. Hence the absorption in black hole back ground is generically non-zero and unlike the free fermion case, where Landau levels are true energy eigen states, any excitation with a generic integral quantum number $n$ and quantum number $\omega$ gets soaked up by the black hole.

The situation changes near a pole of the spectral function and absorption by the black hole goes to zero. At finite $h$, existence  of a pole of the spectral function at $L_n=k_F,\omega=0$ does not imply that $k_F$ is a physical value of $L_n$. However, as the value of $h$ is gradually changed the pole passes through allowed values of $L_n$ (i.e. $k_F(q_{eff})=\sqrt{2nqh}$ for some $n$ and $h=h_n$). In  the limit $h \rightarrow h_n$, the position of the pole of $\mathrm{Im}~G^{(n)}_{22}(\omega)$ on the complex $\omega$ plane approaches $\omega=0$. Just like \eref{pole} and \eref{critical} one may define two critical exponent,
\begin{eqnarray}
 \omega_*(h) &\sim& (h-h_n)^\alpha \\
\nonumber \Gamma_*(h) &\sim& (h-h_n)^\beta.
\end{eqnarray}
It is not difficult to see that $\alpha=z(q_{eff})$ and $\beta=\delta(q_{eff})$, where $q_{eff}=q_{ini}  \sqrt{1-\frac{h_n^2}{3}}$. (see \eref{scaling1}, \eref{scaling2})

Generically the pole moves to smaller values of $n$ with increasing $h$ (\fref{fig:poles_moving}). Due to the gradual passing of poles through physical values of $L_n$, the spectral function diverges periodically with changing magnetic field. Due to the existence of a long living excitation near this point, it is expected that other quantities associated with the system may also show a similar periodic divergence. This is similar to the de Haas-van Alphen effect (see Sec. \ref{condmat}) observed in condensed matter systems.  We can indeed derive a formula similar to Onsager's relation (\eref{period}) in the limit of a small magnetic field.

\subsubsection{Small magnetic field}

The divergence in the imaginary part of the Green's function occurs when the condition $k_F(q_{eff}) = \sqrt{2nqh}$ is satisfied.
For small $h$, $q \approx q_{eff}$ at linear order in $h$. Hence if the condition is satisfied for two adjacent level with quantum number $n$ and $n-1$,
where the magnetic field takes the values $h$ and $h+\delta h$ respectively then
\begin{eqnarray}
& &\sqrt{2nqh}=k_F=\sqrt{2(n-1)q(h+\delta h)} \\
&\Rightarrow& (1 +\frac{1}{n})(h-\delta h)=h \\
&\Rightarrow&  \delta h=\frac{h}{n}.
\end{eqnarray}
Here we have neglected the quadratic terms in both $h$ and $\frac{1}{n}$, as $\frac{1}{n}\sim O(h)$. Now using $n = k_F^2 / 2qh$ we get
\begin{eqnarray}
& & \delta h = \frac{2 q h^2}{ k_F^2} \\
&\Rightarrow&  |\delta(\frac{1}{h})|=\frac{2\pi q}{\pi k_F^2}= \frac{2\pi q}{ A_F},
\end{eqnarray}
where $A_F$ is the area of the Fermi surface. This is our version of Onsager's relation.

\subsubsection{Strong magnetic field}

At strong magnetic field ($h\rightarrow \sqrt{3}$), the first non-trivial level is at $L_1=\sqrt{2 \sqrt{3} q}$. Interesting physical regions seem to lie inside this level. One expects to see something like quantum hall effect etc in this regime. A natural question is whether it is possible to relate the movement of the Fermi level to a change in the filling fractions of the degenerate Landau levels and investigate QHE in this setup\footnote{We thank Hong Liu for discussions on this matter.}. We leave this issue for scrutiny in a future work.

\iffalse
For strong magnetic fields ($h\rightarrow \sqrt{3}$) the first non-trivial level is at $L_1=\sqrt{2 q h}\approx \sqrt{2\sqrt{3} q}$.
Also $q_{eff}$ will be very small. It seems that any interesting physical region will be below the first level.
What physical significance this regime may carry is not clear. One may expect to see something like Quantum Hall effect etc.
This needs further investigation.
\fi

\section{Condensed Matter Physics}\label{condmat}

In this section we shall review some interesting aspects of the relevant condensed matter physics. We are mainly interested in the transport properties of strongly coupled electron systems. For this the most relevant physics comes from low energy excitations near the Fermi surface. We are also interested in the behaviour of the Fermi surface in presence of a uniform magnetic field.  To this end, let us first summarize some well known experimental effects in condensed matter physics \cite{Asmer}.

\subsection*{de Haas-van Alphen Effect}

\begin{figure}[htbp]
\begin{center}
\includegraphics[scale=1]{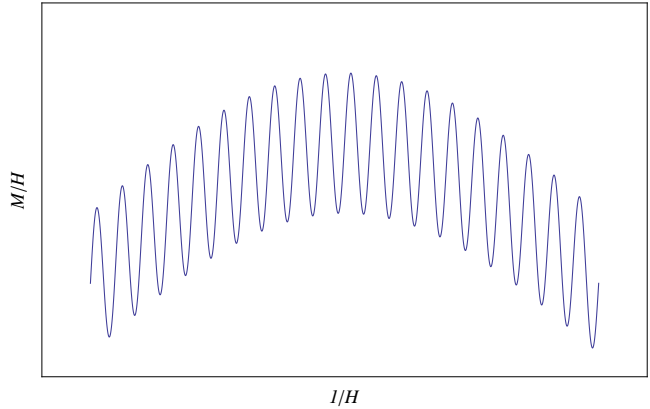}
\caption{Oscillations in the susceptibility with varying magnetic field.}
\label{oscillations}
\end{center}
\end{figure}

This effect was observed originally by de Haas and van Alphen in 1930 for Bismuth and subsequently by others for many different metals. When a metallic sample is placed in a high magnetic field $H$ at low temperatures, many physical parameters oscillate as the magnetic field is varied. For example, the magnetization $M$ to magnetic field ratio $M/H$ shows characteristic oscillations with varying $H$. More precisely, the quantity that is measured is the magnetic susceptibility $\chi = dM/dH$. Fig. \ref{oscillations} shows a schematic plot of the susceptibility illustrating this effect.
In presence of a uniform transverse magnetic field the electron orbits in a metallic sample form quantized Landau levels. For a 2+1 dimensional system of electrons the levels are characterized by a single integer $\nu$. The energy of the level $\nu$ is given by
\begin{equation}
\label{landau}
E_{\nu} = \left( \nu + \frac 12 \right) \hbar \omega_{c}
\end{equation}
where $\omega_{c} = eH/mc$ is the cyclotron frequency, $e$ and $m$ being the electronic charge and mass respectively. The corresponding wavefunctions are Hermite polynomials, similar to a harmonic oscillator. Each Landau level has a very high degeneracy proportional to the magnetic field $H$.

The de Haas-van Alphen effect turns out to be a direct consequence of the above mentioned quantization of  closed electronic orbits in an external magnetic field. The oscillations were first explained by Landau using this picture.  It turns out that the susceptibility is actually a periodic function of the inverse magnetic  field $1/H$. A remarkable observation due to Onsager which connects the oscillations with the properties of the Fermi surface is expressed through the formula
\begin{equation}
\label{period}
  \Delta\left(\frac 1H\right) = \frac{2\pi e}{\hbar c} \frac{1}{A}
\end{equation}
where $A$ is an extremal cross-sectional area of the Fermi surface in a plane normal to the magnetic field. To derive this formula it is assumed that relevant Landau levels have very high level number ($n$). Magnitude of $n$ is given by $E_F/[(e\hbar/mc)H]$, where $E_F$ is the Fermi energy. Typically $E_F/[e\hbar/mc]\sim 10^8 G$. Hence $n \sim 10^4$ even at a field strength of the order $10^4 G$.

Let us now briefly describe the mechanism behind the above oscillatory behaviour. In metals, it is expected that the dynamics near the Fermi surface governs most of the electronic properties. The density of electronic states near the Fermi level is thus an important factor. It can be shown that whenever the area of a quantized electronic orbit coincides with an extremal cross sectional area of Fermi surface in a plane normal to the magnetic field there is a singularity of the density of states at the Fermi level. As the magnetic field is varied the successive quantized orbits cross this extremal area. This is responsible for the observed oscillatory behaviour.

The de Haas-van Alphen effect is thus a very useful tool for probing the Fermi surface.

In addition to the de Haas-van Alphen effect, some related phenomena exist for metals which can also provide some information about the Fermi surface. Some examples are the magnetoacoustic effect, ultrasonic attenuation, the anomalous skin effect,  cyclotron resonance, size effects, etc.

\subsubsection*{Quantum Hall Effect}

The conventional Hall effect is the phenomenon where a conductor carrying an electrical current when placed in a transverse magnetic field develops a voltage across its edges transverse to both the magnetic field and the current. Classically, the effect is easily explained as resulting from the deflection of the electrons carrying the current due to the Lorentz force in presence of the magnetic field.

This picture changes as impurities are introduced into the system. This leads to localized bound states with energies between the Landau levels, at the expense of some of the degenerate Landau level states. Now as the electron density is increased beyond a filled Landau level, the extra electrons fill the localized states. The Hall conductivity is unchanged during this process as the localized states carry no current. Only after enough electrons have been added to raise the Fermi energy to lie in the next highest Landau level does the conductivity begin to increase again. The process repeats itself, which leads to plateaus in the Hall conductivity. This phenomenon is known as the \emph{integer} quantum Hall effect, as the plateaus occur after each Landau level is completely filled. The integer quantum Hall effect dominates when the electrons are weakly interacting; in presence of strong electron-electron interactions Hall plateaus can occur when only a fraction of the states at a particular Landau level is filled. This is known as the \emph{fractional} quantum Hall effect. These effects are prominent at low temperature and high magnetic field. It would be interesting to investigate such effects in our system.

\section{Conclusions and Future Directions}
\label{conclusions}

\nt In this paper we have taken the first steps towards understanding the effects of an external magnetic field on the Fermi level structure of a strongly coupled electronic system from the holographic viewpoint. The results are quite interesting. It turns out that the holographic system with the magnetic field can be simulated by just changing the charge $q$ of the fermion field $\Psi$ (\eref{rel}). Thus all of our results can be mapped back to the zero field case. From this observation, we address phenomena like gradual disappearance of Fermi surface at large magnetic field etc. We also see quantization of the levels similar to Landau levels arising in ordinary metals due to quantization of electron orbits.  Consequently there exists a periodic divergence in the spectral function with changing magnetic field; this behaviour is expected to persist in other physical parameters which can be measured. It is expected that at finite temperature the divergence will be replaced by finite peaks. This is reminiscent of the de-Haas van Alphen effect seen in metals. So the holographic non-Fermi liquid seems to possess some of the properties expected in a condensed matter system in a magnetic field. In light of the methods introduced in \cite{Faulkner:2009wj} it is possible to analyze the system analytically. This will provide a better understanding of the behaviour of Fermi surfaces in presence of intermediate and strong magnetic fields.

There are some interesting physical questions which we plan to investigate in future projects. One issue concerns the behaviour of the system in global $AdS$ space, which is equivalent to putting the holographic CFT in a finite volume. This provides another parameter which we can tune to observe its effects on the system. This might allow better control of the filling fraction of the degenerate states within a Landau level and thus facilitate the study of the QHE. We would also like to consider a situation where there is a condensate of a composite particle made up of the fermions (similar to the holographic superconductor) and investigate what happens to the Fermi level as the condensate turns on. Recently an extremal geometry with a non-zero condensate is found \cite{Gubser:2009qm}, this may be important for our purpose. Another topic which is interesting is the quantum Hall effect. There has recently been some work in this direction \cite{Davis:2008nv,KeskiVakkuri:2008eb}, and we would like to further investigate similar phenomenon.

\section{Acknowledgements}
We thank Hong Liu, Mark van Raamsdonk and Moshe Rozali for helpful suggestions and comments on our work. We would also like to thank the String Theory Group at UBC for their support and encouragement. PB and AM acknowledge support from the Natural Sciences and Engineering Research Council of Canada.

\bibliography{fermion2.bib}
\end{document}